\documentclass[journal]{IEEEtran}

\ifCLASSINFOpdf

\else

\fi

\usepackage[utf8]{inputenc}
\usepackage{amsmath}
\usepackage{float}
\usepackage{mathtools}
\usepackage{braket}
\usepackage{graphicx}
\graphicspath{{./figures/}}
\usepackage{amsfonts}
\usepackage[ruled,vlined,linesnumbered]{algorithm2e}
\let\oldnl\nl% Store \nl in \oldnl
\newcommand{\nonl}{\renewcommand{\nl}{\let\nl\oldnl}}% Remove line number for one line

\usepackage{tikz}
\usetikzlibrary{shapes.geometric, arrows}

\usepackage{csquotes}
\usepackage{hhline}
\usepackage{amssymb}
\usepackage{listings}
\usepackage{color}
\definecolor{codegreen}{rgb}{0,0.6,0}
\definecolor{codegray}{rgb}{0.5,0.5,0.5}
\definecolor{codepurple}{rgb}{0.58,0,0.82}
\definecolor{backcolour}{rgb}{0.95,0.95,0.92}
\usepackage{soul}
\usepackage{graphicx}   % For other  mats (e.g., PNG, JPG)
\usepackage{svg} 
\usepackage{subcaption}
\usepackage{titlesec}

\usepackage{dcolumn}
\usepackage{tabularx}
\usepackage{hyperref}
\hypersetup{
    colorlinks=true,
    linktoc=all,
    linkcolor=black,
    citecolor=magenta,
}
\usepackage{longtable}
\usepackage{braket}
\newcolumntype{C}{>{\centering\arraybackslash}X}
\SetAlFnt{\small}
\usepackage[font=small]{caption}
\soulregister{\cite}{1} % Register \cite with soul
\soulregister{\ref}{1}  % If using \ref inside \hl

\tikzstyle{startstop} = [rectangle, rounded corners, minimum width=3cm, minimum height=1cm,text centered, draw=black, fill=red!30]
\tikzstyle{process} = [rectangle, minimum width=3cm, minimum height=1cm, text centered, draw=black, fill=blue!30]
\tikzstyle{decision} = [diamond, minimum width=3cm, minimum height=1cm, text centered, draw=black, fill=green!30]
\tikzstyle{arrow} = [thick,->,>=stealth]

\begin{document}
% correct bad hyphenation here
\hyphenation{op-tical net-works semi-conduc-tor}

\title{QDCNN: Quantum Deep Learning for Enhancing Safety and Reliability in Autonomous Transportation Systems}

\author{Ashtakala Meghanath, Subham~Das, Bikash~K.~Behera, Muhammad Attique Khan, Saif~Al-Kuwari and Ahmed~Farouk
\thanks{Ashtakala~Meghnath is with the Department of Physics, Indian Institute of Science Education and Research, Thiruvananthapuram, India;  Email: meghanath19@alumni.iisertvm.ac.in}
\thanks{Subham Das is with the Department of Physics, Indian Institute of Science Education and Research, Thiruvananthapuram, India; Email: Subhamdas19@alumni.iisertvm.ac.in}
\thanks{$^{*}$ have equal contributions to the manuscript}
\thanks{B.~K. Behera is with Bikash's Quantum (OPC) Pvt. Ltd., Mohanpur, WB, 741246 India; Email: bikas.riki@gmail.com}
\thanks{Muhammad Attique Khan is with the Department of AI, College of Computer Engineering and Science, Prince Mohammad Bin Fahd University, Al Khobar, Saudi Arabia; Email: attique.khan@ieee.org}
\thanks{S.~Al-Kuwari is with the Qatar Center for Quantum Computing, College of Science and Engineering, Hamad Bin Khalifa University, Doha, Qatar. e-mail: (smalkuwari@hbku.edu.qa).}
\thanks{A.~Farouk is with the Qatar Center for Quantum Computing, College of Science and Engineering, Hamad Bin Khalifa University, Doha, Qatar, and with the Department of Computer Science, Faculty of Computers and Artificial Intelligence, Hurghada University, Hurghada, Egypt; Email: ahmed.farouk@sci.svu.edu.eg}

\thanks{\textit{Corresponding Authors: Ahmed~Farouk}}
}

\maketitle
\begin{abstract}
%edited
In transportation cyber-physical systems (CPS), ensuring safety and reliability in real-time decision-making is essential for successfully deploying autonomous vehicles and intelligent transportation networks. However, these systems face significant challenges, such as computational complexity and the ability to handle ambiguous inputs like shadows in complex environments. This paper introduces a Quantum Deep Convolutional Neural Network (QDCNN) designed to enhance the safety and reliability of CPS in transportation by leveraging quantum algorithms. At the core of QDCNN is the UU† method, which is utilized to improve shadow detection through a propagation algorithm that trains the centroid value with preprocessing and postprocessing operations to classify shadow regions in images accurately. The proposed QDCNN is evaluated on three datasets on normal conditions and one road affected by rain to test its robustness. It outperforms existing methods in terms of computational efficiency, achieving a shadow detection time of just 0.0049352 seconds, faster than classical algorithms like intensity-based thresholding (0.03 seconds), chromaticity-based shadow detection (1.47 seconds), and local binary pattern techniques (2.05 seconds). This remarkable speed, superior accuracy, and noise resilience demonstrate QDCNN’s —key factors for safe navigation in autonomous transportation in real-time. This research demonstrates the potential of quantum-enhanced models in addressing critical limitations of classical methods, contributing to more dependable and robust autonomous transportation systems within the CPS framework.
\end{abstract}


\IEEEpeerreviewmaketitle

\section{Introduction}\label{QVP:Sec1}
Autonomous transportation has emerged as a revolutionary technology in different forms, such as self-driving cars, autonomous trains, and smart traffic management systems. It addresses key transportation challenges, including human error, road accidents, traffic congestion, and environmental sustainability \cite{liu20246g}. By executing precise decision-making, autonomous systems can reduce the likelihood of collisions \cite{que_Wang2024}. To achieve these goals, integrating CPS in transportation is essential \cite{9695482}. CPS refers to systems where physical infrastructure and computational systems interact in real time, enhancing efficiency, safety, and reliability. However, despite their increasing demand and importance, autonomous transportation systems still face significant challenges, such as computational complexity, the need for large-scale real-time data processing, and the development of reliable decision-making algorithms to ensure safety in unpredictable environments \cite{khalil2024advanced}. Recently, human-inspired methods derived from cognitive science and neuroscience have been explored to address these challenges \cite{PLEBE2024101169}. Furthermore, a multi-agent reinforcement learning framework designed for biometric ticketing in multi-transport environments is proposed in \cite{LAKHAN2024102471}.  A secure blockchain-based framework for mitigating cyber-attacks in V2I systems is presented in \cite{LAKHAN2024111576}. Meanwhile, a distributed hybrid decision-making framework for coordinating multiple autonomous vehicles on multi-lane highways is presented in \cite{fabiani2019multi}. A vehicle-pedestrian negotiation model that enhances traffic flow by facilitating the exchange of cues between vehicles and pedestrians is proposed in \cite{aradi2020survey}. In \cite{gupta2018negotiation}, a vehicle motion strategy is developed to simulate real-world negotiation scenarios, reducing travel times compared to the common practice of always stopping for pedestrians. A decision-making framework for autonomous vehicles using a partially observable Markov decision process, which predicts the intentions of other vehicles based on noisy sensor data, is introduced in \cite{hubmann2018automated}.
Deep Neural Networks (DNNs) are essential in autonomous transportation, performing critical tasks such as decision-making, object detection \cite{song2024robustness}, identifying other vehicles \cite{lin2024clothoid}, traffic sign recognition \cite{649946}, forecasting Norwegian air passenger traffic \cite{stanulov2023comparative} and shadow detection \cite{10.1007/978-3-031-48751-4_16}.
\begin{figure*}
\centering
\includegraphics[width=\linewidth]{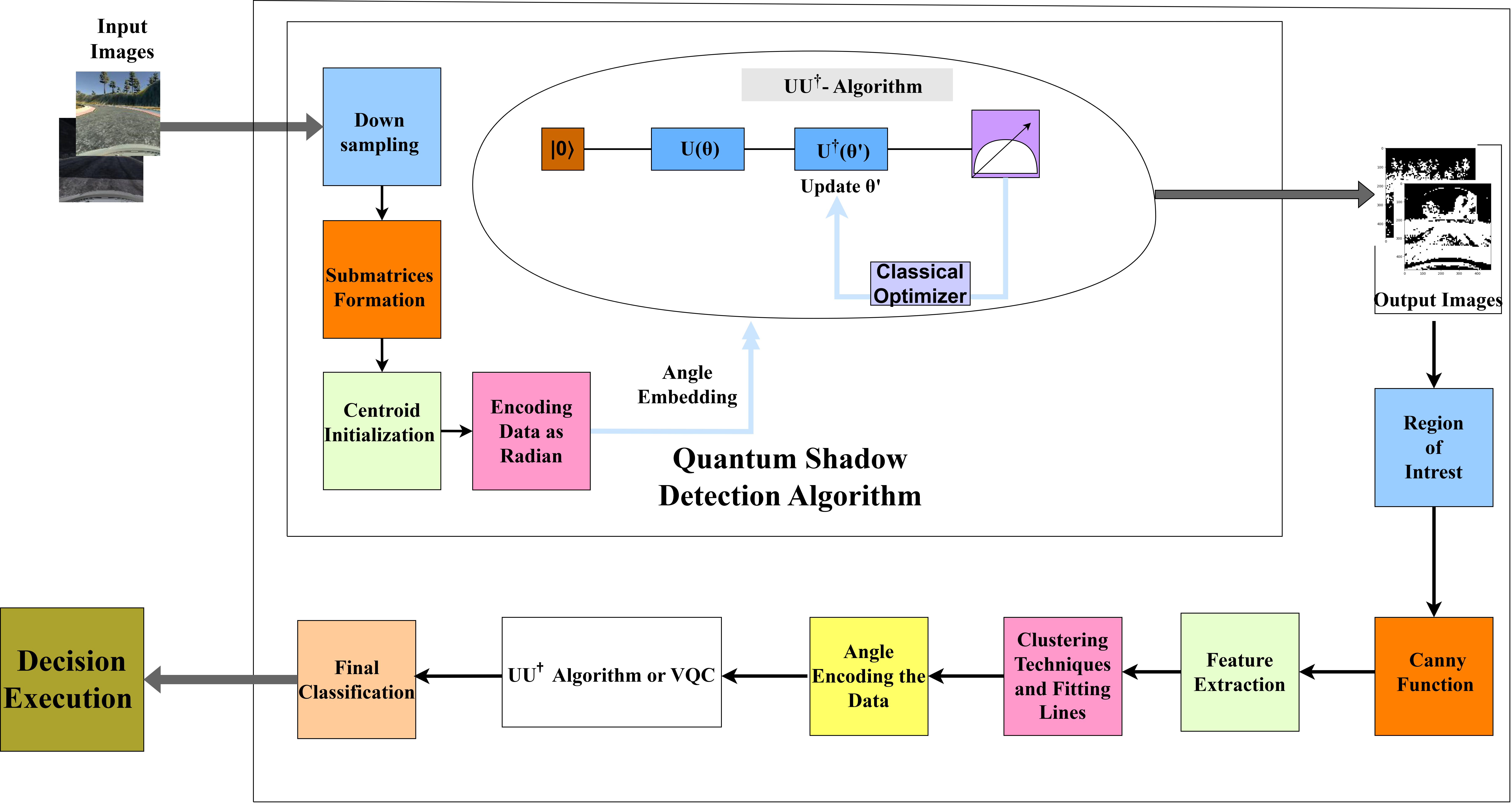}
\caption{QDCNN System Model.}
\label{fig_1}
\end{figure*}

%\hl{The industrial landscape for autonomous transportation is rapidly evolving, with significant contributions from global leaders such as Tesla, Waymo, Uber, and Nvidia. These companies are driving innovation in self-driving cars, logistics, and smart city transportation systems, aiming to enhance safety, efficiency, and environmental sustainability. While advancements in classical and quantum computing have significantly contributed to autonomous transportation, there remains a pressing need for solutions capable of addressing the challenges posed by real-world scenarios.} 
Shadow detection is a critical challenge in autonomous driving, as shadows can distort the input to DNNs, causing inaccurate predictions. %Therefore, many studies have developed classical algorithms for detecting shadows in various situations, aiming to improve the accuracy and reliability of autonomous systems.
However, the proposed systems face challenges, particularly in image processing, which is crucial for safe navigation. Issues like shadows, lighting, and object detection precision can disrupt DNN performance, leading to incorrect predictions and fatal accidents \cite{app13169313}. Other challenges include scalability, adapting to new environments, path planning \cite{que_Chen2024}, communication constraints, security, and fault tolerance\cite{song2024robustness}. While approaches like conditional imitation learning have progressed, they still struggle in unseen environments, varying weather conditions, and avoiding static obstacles \cite{9928072}. 

Quantum computing (QC) offers significant potential to address computational challenges in autonomous transportation \cite{zhuang2024quantumcomputingintelligenttransportation}. Quantum algorithms, such as the UU$^{\dagger}$ and Variational Quantum Classifier (VQC) \cite{azad2022solving}, can process large datasets and perform complex calculations exponentially faster than classical algorithms, improving decision-making in real-time.  For example, a quantum-inspired lattice Boltzmann model was introduced to enhance the modelling of pedestrians' often irrational behaviour in traffic scenarios, offering improvements over classical models \cite{9514552}. Furthermore, a quantum game theory is employed to challenge the traditional assumption of rationality in decision theory, aiming to improve the realism of decision-making in autonomous driving systems when interacting with humans \cite{Song20223}. A Quantum Neural Network (QNN) framework to effectively process and solve the challenge in current vehicle road cooperation systems is proposed in \cite{innan2024qnn}.
Despite these advancements, there are still gaps in the current literature, particularly in ensuring the safety and reliability of autonomous systems. Classical models often struggle with real-time decision-making under uncertain conditions caused by shadows, leading to potential safety risks \cite{wevj15030099}. Quantum algorithms show promise in improving speed and efficiency. Still, limited research has examined their ability to handle shadow-related errors and edge cases in real-world autonomous transportation. Therefore, we proposed a new QDCNN algorithm (Fig. \ref{fig_1}) to directly address the challenges associated with shadow detection in autonomous transportation systems, which are difficult for both classical and existing quantum models to handle effectively. Furthermore, it is extended to a quantum model that predicts the direction of motion by using lane detection and requiring one qubit with two gates, which is more resource efficient compared to other quantum thresholding algorithms \cite{barui2023novelapproachthresholdquantum}. The QDCNN uses the UU$^{\dagger}$ method to enhance the detection of shadows in real-time, improving the system's ability to identify shadowed regions in the Region of Interest (ROI) of self-driving cars using Intensity and chromaticity-based techniques. To our knowledge, this is the first study to combine QC and deep learning (DL) for shadow detection. Furthermore, the existing algorithms have not been verified in the presence of different noisy channels for robustness.

The contributions of this paper can be summarized as follows:
\begin{enumerate}
\item A resource and time-efficient novel QDCNN model is proposed, utilizing UU$^{\dagger}$ for image processing, shadow detection, and improving decision-making in self-driving cars.
\item The QDCNN is applied to threshold images, classifying them into black and white images using only one qubit and two gates with comparatively minimum quantum resources compared to existing models.
\item The QDCNN model's superiority in processing time is demonstrated through performance comparisons with classical models using gate operation time of qubits. Additionally, its reduced resource requirements make it space-efficient, presenting a promising solution to the space constraints in self-driving cars.
%\item Within the QDCNN, a simple decision-making model is proposed based on the input that can make simple yet critical decisions about whether the road ahead is left, right or straight. 
\item The robustness of QDCNN is verified against six different noise models, highlighting its resilience to noise, which is crucial for ensuring safety and reliability in autonomous transportation systems. 
\end{enumerate}
The rest of the paper is organized as follows: Section \ref{SecII} covers the methodology and algorithms used. Section \ref{SecIII} presents the results of applying the proposed QDCNN on the given datasets. Finally, Section \ref{SecIV} discusses the results, offering concluding remarks and suggestions for future work.

\section{Methodology \label{SecII}}
The development process of shadow and road detection follows a structured sequence of stages as illustrated in Fig \ref{fig_1}. Initially, the image undergoes preprocessing, beginning with shadow removal using a QDCNN to detect shadowed regions. The image is then converted to grayscale to simplify subsequent operations like edge detection using tools from the OpenCV package like Gaussian blur and the Canny function. Next, an ROI is identified to exclude irrelevant areas, and the Hough transformation is applied to detect road lanes by identifying straight lines within the ROI. However, due to the dependence of Hough transformation on various parameters, the result often includes segments of lines rather than continuous ones. Techniques are applied to extract critical parameters such as the number of lines, slopes, and intercepts. In our proposed model,  two slopes were used, and the dimension of the input is $2 \times 1$ vector. Finally, two algorithms, the $UU^{\dagger}$ method and the VQC use these parameters for decision-making and final predicting regarding directional inclination, which is crucial for autonomous navigation. 
\subsection{Preprocessing}
In the self-driving car's lane detection process, image preprocessing begins with identifying and replacing shadow regions with road pixel values to create a consistent surface (Algorithm \ref{algo_lane_detection}). Edge detection using the Canny function follows, focusing on a designated ROI that excludes unnecessary areas. The Hough line transformation is then applied to detect straight lines, representing road lanes. The effectiveness of this process depends on correctly tuning parameters like edge thresholds and line properties, ensuring accurate lane identification for safe vehicle navigation.
\subsubsection{Training for the Centroid Value}
The proposed propagation algorithm for training the centroid value of the activation function, using the \( U U^\dagger \)\ algorithm, involves the following steps: First, crop the training region of the image, and then either guess an initial centroid value or calculate one through a clustering method. The feature values are converted into radians, and a corresponding rotation matrix is created. By multiplying two rotation matrices, the angle of the resulting product matrix (\( \tau \)) is determined as shown in Eq. \ref{abceqn}, where D is the matrix trace.
\begin{eqnarray}
D&=& 2\times\cos{\tau}\\
\frac{dD}{d\theta}&=& -2\times\sin{\tau}
\label{abceqn}
\end{eqnarray}
The trace and its derivative are calculated, and the centroid is found by using the inverse of the derivative in a recursive formula (Eqs. 3-4).
\begin{align}
    \theta_{n+1} - \theta_{n} &= \frac{d\theta}{dD} \times \delta D
\end{align}
Here, $\theta$ is the parameter that is trained, and the value of $D$ corresponds to
\begin{align}
    \delta D=2\cos{\tau_{n}}-2
\end{align}
The final centroid value obtained from the last iteration is the trained centroid for the algorithm.
\subsubsection{Shadow Detection}
The shadow detection algorithm downscales the image for faster processing and divides it into \(n \times n\) submatrices. It can be implemented using two types of input-Intensity or chromaticity. For the intensity-based method, the algorithm uses greyscale images to calculate the intensity for detecting contours and is less affected by conditions such as fog, glare, or shadow compared to color images, making them suitable for preprocessing or fallback strategies. For the chromaticity-based technique, the chromaticity is calculated directly from the RGB image. The midpoints of the submatrices are used as input for the \( U U^\dagger \) method to classify regions as shadow or shadowless based on centroid values. Shadow regions are refined using a propagation algorithm, assigned a value of 255, while shadowless regions are set to 0. A median filter is applied to enhance the image, and the algorithm is tested on sample images to evaluate its performance and effectiveness.
%The shadow detection algorithm downscales the image for faster processing and divides it into \(n \times n\) submatrices. \hl {A Quantum Binary Image Converter (QBIC) algorithm proposed for generating black and white images from greyscale images for detecting contours and is less affected by conditions such as fog, glare, or shadow compared to colour images, making them suitable for preprocessing or fallback strategies. For QBIC, midpoints of the submatrices are used as input for the} \( U U^\dagger \) \hl{method to classify regions as shadow or shadowless based on centroid values. Shadow regions are refined using a propagation algorithm, assigned a value of 255, while shadowless regions are set to 0. A median filter is applied to enhance the image, and the algorithm is tested on sample images to evaluate its performance and effectiveness.}

\subsubsection{Road Detection}
The lane detection process, outlined in Algorithm \ref{algo_lane_detection}, involves several key preprocessing steps using OpenCV. It begins with shadow removal, followed by applying a median filter and Gaussian blur to reduce noise and enhance clarity for edge detection. The Canny function highlights sharp intensity changes, and an ROI is defined to focus on relevant parts of the image. Road lanes are then detected using the Hough transformation. To simplify and organize the detected lines, three clustering methods—image-split, k-means \cite{k-means}, and spectral clustering \cite{spectral-clustering}—are applied. Each method groups the lines and generates representative slopes, which serve as inputs for the final predictions using quantum hybrid algorithms such as the \(UU^{\dagger}\) algorithm or VQC, enhancing the accuracy of lane detection.
\begin{algorithm}
\DontPrintSemicolon
\caption{$UU^\dagger$ Method}
\label{UU algo}

\nonl\textbf{Input:}  Normalized cleaned and scaled dataset\;
\nonl\textbf{Output:} accuracy \;
    
\SetKwFunction{CircuitFunction}{Circuit\_function}
\SetKwProg{Fn}{def}{:}{end}
\Fn{\CircuitFunction{$params$}}{
    \nonl\textbf{Apply unitary gates}\;
    \nonl    Initialize quantum state $\ket{0}$\;
\nonl    Apply unitary gate $U(\theta)$ to the quantum state: $\ket{\psi} \gets U(\theta) \ket{0}$\;
\nonl    Apply unitary gate $U(-\theta')$ to the quantum state: $\ket{\psi'} \gets U(-\theta') \ket{\psi}$\;
    \textbf{Perform Measurement}\;
\nonl    Measure the circuit in the computational basis.\;

    \textbf{Calculate inner product}\;
\nonl    Compute the inner product between the two measurement outcomes by taking the square of the probability of $\ket{0}$\;

 \textbf{Classification}\;
    
\nonl  The classification condition on the inner product is applied\;
    \textbf{accuracy}\;
\nonl The accuracy is calculated 
    
}

\end{algorithm}

\subsection{\texorpdfstring{$UU^{\dagger}$}{UU-dagger} Algorithm}

%\subsection{$UU^{\dagger}$ Algorithm}
The $UU^{\dagger}$ algorithm calculates the inner product between the centroid and test data encoded in quantum states as given in Algorithm \ref{UU algo} and Eqs. 5-9. First, the centroid data is encoded into a quantum state \(|C\rangle\) using the unitary operator \( U_1 \), 
\begin{equation}
|C\rangle = U_1 |0\rangle^{\otimes n}
\end{equation}
After that, another unitary operator \( U_2 \) is used to encode the test data into a quantum state \(|T\rangle\),
\begin{equation}
|T\rangle = U_2 |0\rangle^{\otimes n}
\end{equation}
Finally, the inner product between \( |C\rangle \) and \( |T\rangle \) can be calculated as the square root of the probability of measuring the state \(|0\rangle^{\otimes n}\) after applying the operator \(A\) to it.
\begin{eqnarray}
A &=& U_2^\dagger U_1,\\
\langle C | T \rangle &=& \langle T | C \rangle = ^{n \otimes}\braket{0|A|0}^{\otimes n},\\
\langle C | T \rangle &=& \sqrt{P_{|0\rangle^{\otimes n}}} 
\end{eqnarray}
After applying the Hough transformation, which produces multiple lines, the goal is to consolidate these lines into two representative lines for further analysis. This is achieved through three techniques: the image-split method, k-means clustering, and spectral clustering. These methods allow the \(UU^{\dagger}\) algorithm to effectively reduce the multiple detected lines into one or a few representative lines based on their slopes. This consolidation enhances the accuracy and reliability of decision-making processes. %The step-by-step process of \(UU^{\dagger}\) algorithm is given in Algorithm \ref{UU algo}.%The circuit for the \(UU^{\dagger}\) method is illustrated in Figure \ref{UU dagger diagram}.

\begin{algorithm}
\DontPrintSemicolon
\SetKwInOut{Input}{Input}
\SetKwInOut{Output}{Output}

\Input{Image Path}
\Output{Direction of motion}

\SetKwFunction{ROI}{region\_of\_interest}
\SetKwFunction{ShadowDetect}{Shadow\_detection}
\SetKwFunction{Canny}{Canny}
\SetKwProg{Fn}{def}{:}{end}

\Fn{\ROI{image}}{
    \nonl Determine the dimensions of the image (height, width)\;
    \nonl Define quadrilateral points to create a mask\;
    \nonl Fill the mask with white colour and apply it to the image\;
    \Return masked image\;
}

\Fn{\ShadowDetect{image}}{
    \nonl Downsample the image and divide it into $n \times n$ submatrices\;
    \nonl Find the average value of the centre submatrices\;
    \nonl Use the average as input for $UU^{\dagger}$ to classify shadow regions\;
    \nonl Assign values: shadow (255) or shadowless (0) for each submatrix\;
    \nonl Recombine the submatrices into the full image\;
    \Return shadow-detected image\;
}

\BlankLine
Load the image\;
Apply \textbf{Shadow detection} to remove shadows using $UU^{\dagger}$\;
Smooth the image with a \textbf{median filter}\;
Use \textbf{ROI} to detect the road area\;
Replace shadow regions within the ROI with road pixels\;
Apply \textbf{Gaussian blur} and \textbf{Canny edge detection} to enhance edges\;
Crop the image using a second \textbf{ROI}\;
Use \textbf{Hough transform} to detect lines\;
Optimize the lines based on their slope\;
Make a decision on direction using $UU^{\dagger}$ or a VQC model based on the fitted lines\;

\BlankLine
\caption{Lane Detection with Shadow Correction and Edge Detection}
\label{algo_lane_detection}
\end{algorithm}

\begin{figure*}
\centering
\includegraphics[width=\linewidth]{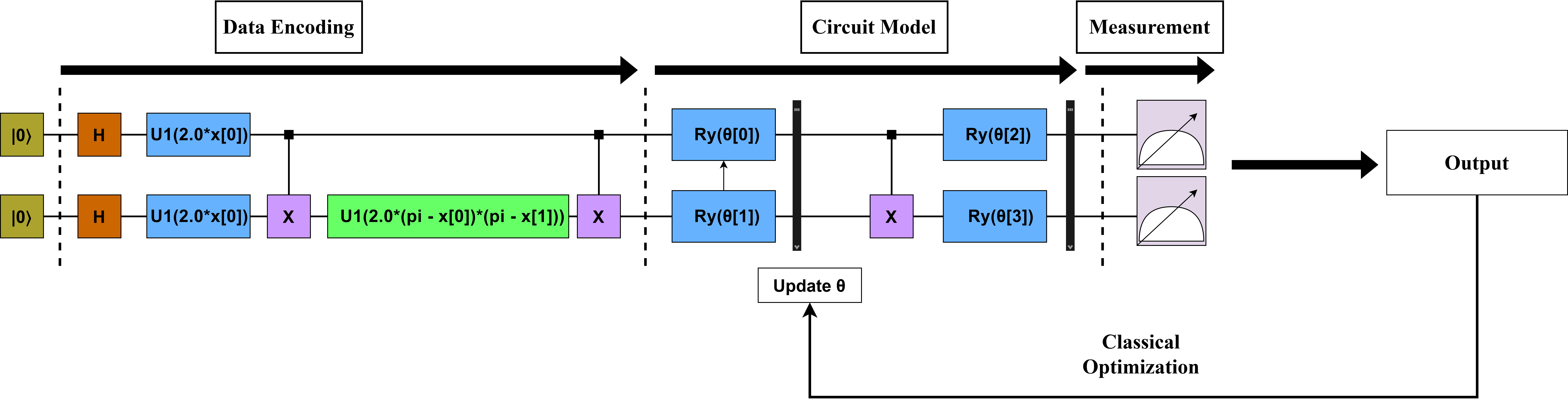}
\caption{VQC Circuit.}
\label{vqc circuit}
\end{figure*}
\begin{algorithm}
\DontPrintSemicolon
\caption{VQC Method}
\label{vqc_callback_algorithm}

\SetKwInOut{Input}{Input}
\SetKwInOut{Output}{Output}

\Input{Slope values list (\texttt{slope\_values\_list}), converted directions (\texttt{converted\_dir})}
\Output{Classification score and plot of objective function values against iteration}

\SetKwFunction{VQC}{VQC}
\SetKwFunction{CallbackGraph}{callback\_graph}
\SetKwFunction{Fit}{fit}
\SetKwFunction{Score}{score}
\SetKwProg{Fn}{def}{:}{end}

\BlankLine
\Fn{\CallbackGraph{weights, obj\_func\_eval}}{
    Append current objective function evaluation to \texttt{objective\_func\_vals}\;
    Clear previous output\;
    Plot objective function value against iteration\;
}

\BlankLine
\textbf{Initialize and configure the VQC model:} \\
\Fn{\VQC{num\_qubits=2}}{
    Instantiate the Variational Quantum Circuit with 2 qubits\;
    Use optimizer \texttt{COBYLA(maxiter=30)} for training\;
    Pass the \texttt{CallbackGraph} function to monitor training progress\;
}

\BlankLine
\textbf{Fit the VQC model to the input data:} \\
\Fn{\Fit{slope\_values\_list, converted\_dir}}{
    Train the VQC model with the provided data\;
    Trigger the \texttt{CallbackGraph} function at each iteration\;
}

\BlankLine
\textbf{Evaluate the VQC model's performance:} \\
\Fn{\Score{slope\_values\_list, converted\_dir}}{
    Calculate the classification score using the input data\;
    \Return Classification score\;
}

\end{algorithm}

\subsection{Variational Quantum Classifier}
The VQC offers a quantum-based alternative to traditional artificial neural networks (ANNs) for classification tasks \cite{jager2023universal}.  After image preprocessing and lane detection, clustering methods such as spectral clustering, k-means clustering, and image splitting are used to group detected lines into two clusters. Linear polynomials are then fitted to each cluster, and the data from these lines is fed into the VQC for final predictions. The COBYLA optimizer is utilized to refine prediction values, while a callback graph function is developed to visualize the optimization process (Algorithm \ref{vqc_callback_algorithm} and Fig. \ref{vqc circuit}).The cost function used to train the model is Mean Squared Error as in Eq. 10., where $y_i$ is the actual value, and $\hat{y}_i$ is the predicted value.
\begin{equation}
\text{MSE} = \frac{1}{N} \sum_{i=1}^{N} (y_i - \hat{y}_i)^2
\end{equation}

%The VQC presents a quantum alternative to conventional artificial neural networks (ANNs) for classification tasks \cite{qiskit2018quantum}. After the preprocessing techniques and lane detection are applied to the images, various clustering techniques, including spectral clustering, k-means clustering, and image splitting, are implemented to form two distinct clusters. For the final prediction, linear polynomials are fitted to each cluster. Subsequently, the data obtained from these fitted lines is input for the VQC \ref{vqc_callback_algorithm}. The COBYLA optimizer is employed to derive the final prediction values. Additionally, a callback graph function is developed to visualize the optimization process. The circuit diagram for VQC is given in figure \ref{vqc circuit}.
\begin{figure*}
\centering
\begin{subfigure}{.28\textwidth}
    \includegraphics[width=\linewidth]{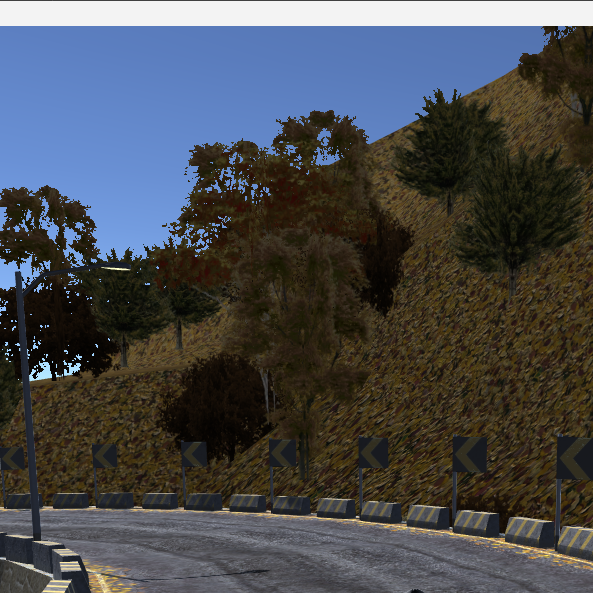 }
    \caption{}
    \label{subfig1a}
\end{subfigure}\hfill
\begin{subfigure}{.28\textwidth}
    \includegraphics[width=\linewidth]{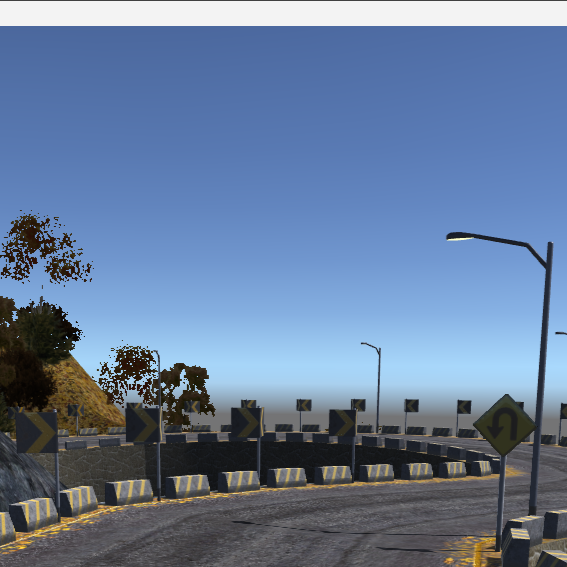}
    \caption{}
    \label{subfig1b}
\end{subfigure}\hfill
\begin{subfigure}{.28\textwidth}
     \includegraphics[width=\linewidth]{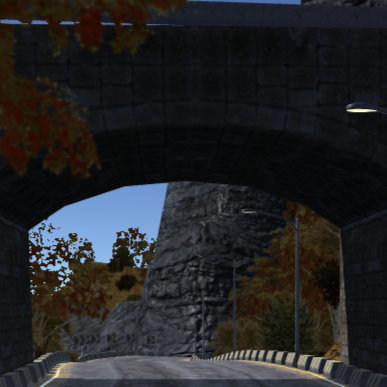}
    \caption{}
    \label{subfig1c}
\end{subfigure}\hfill
\begin{subfigure}{.28\textwidth}
    \includegraphics[width=\linewidth]{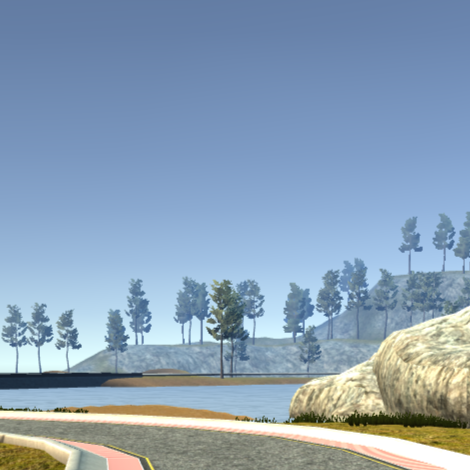 }
    \caption{}
    \label{subfig1a_d2}
\end{subfigure}\hfill
\begin{subfigure}{.28\textwidth}
    \includegraphics[width=\linewidth]{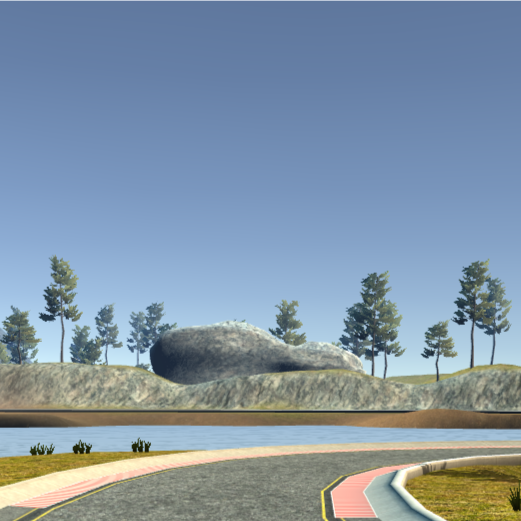}
    \caption{}
    \label{subfig1b_d2}
\end{subfigure}\hfill
\begin{subfigure}{.28\textwidth}
     \includegraphics[width=\linewidth]{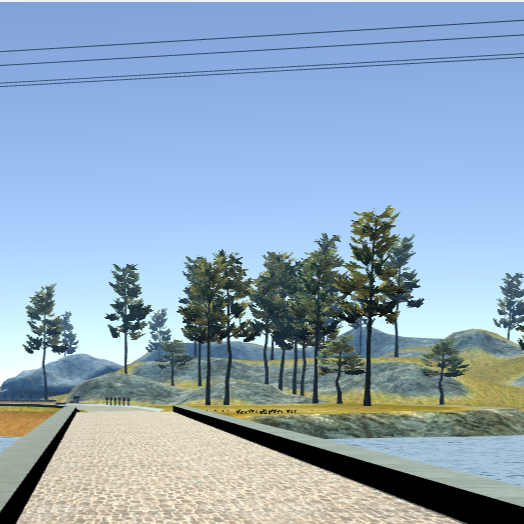}
    \caption{}
    \label{subfig1c_d2}
\end{subfigure}\hfill
\begin{subfigure}{.28\textwidth}
    \includegraphics[width=\linewidth]{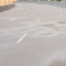 }
    \caption{}
    \label{subfigcarla1}
\end{subfigure}\hfill
\begin{subfigure}{.28\textwidth}
    \includegraphics[width=\linewidth]{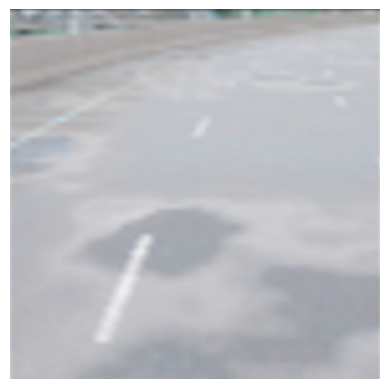}
    \caption{}
    \label{subfigcarla2}
\end{subfigure}\hfill
\begin{subfigure}{.28\textwidth}
     \includegraphics[width=\linewidth]{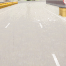}
    \caption{}
    \label{subfigcarla3}
\end{subfigure}\hfill
\caption{Sample Data for the Classification Model. Track-1: (a) Left, (b) Right, and (c) Straight, Track-2: (d) Left (e) Right, and (f) Straight, CARLA (g) Left, (h) Right, and (I) Straight.}
\label{fig4:sampledataformodel}  
\end{figure*}

\begin{figure*}
\centering
\begin{subfigure}{.32\textwidth}
    \includegraphics[width=\linewidth]{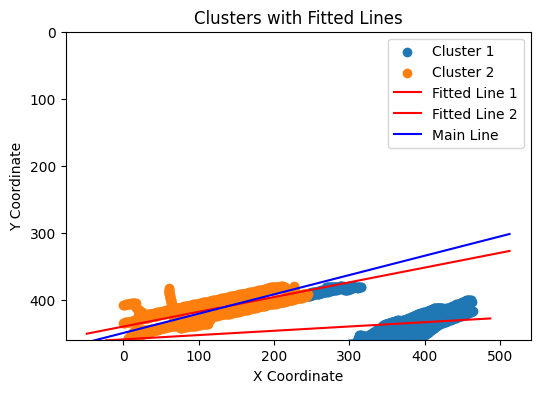 }
    \caption{}
    \label{subfig1a_t1}
\end{subfigure}\hfill
\begin{subfigure}{.32\textwidth}
    \includegraphics[width=\linewidth]{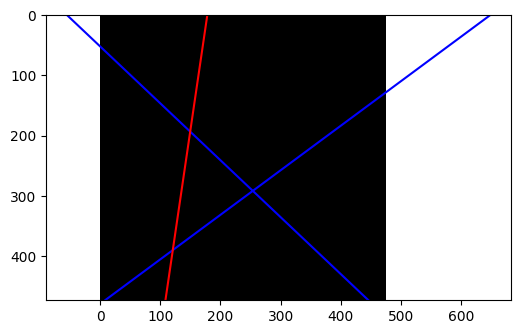}
    \caption{}
    \label{subfig1b_t1}
\end{subfigure}\hfill
\begin{subfigure}{.32\textwidth}
     \includegraphics[width=\linewidth]{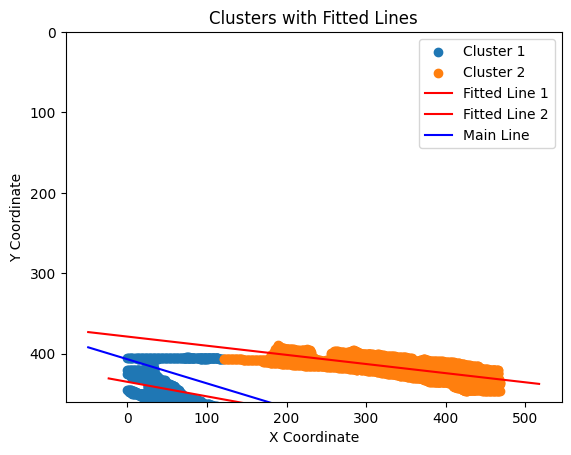}
    \caption{}
    \label{subfig1c_t1}
\end{subfigure}\hfill
\begin{subfigure}{.32\textwidth}
    \includegraphics[width=\linewidth]{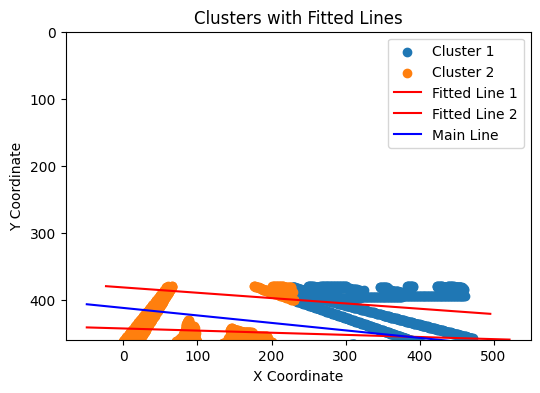}
    \caption{}
    \label{subfig1a_t2}
\end{subfigure}\hfill
\begin{subfigure}{.32\textwidth}
    \includegraphics[width=\linewidth]{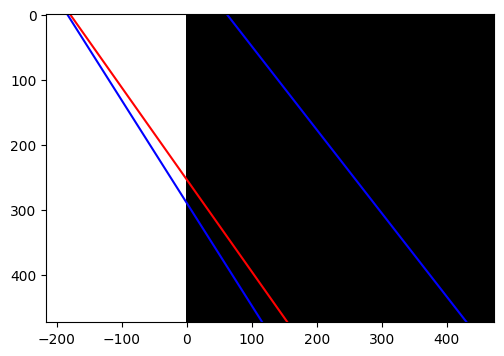}
    \caption{}
    \label{subfig1b_t2}
\end{subfigure}\hfill
\begin{subfigure}{.35\textwidth}
     \includegraphics[width=1.\linewidth]{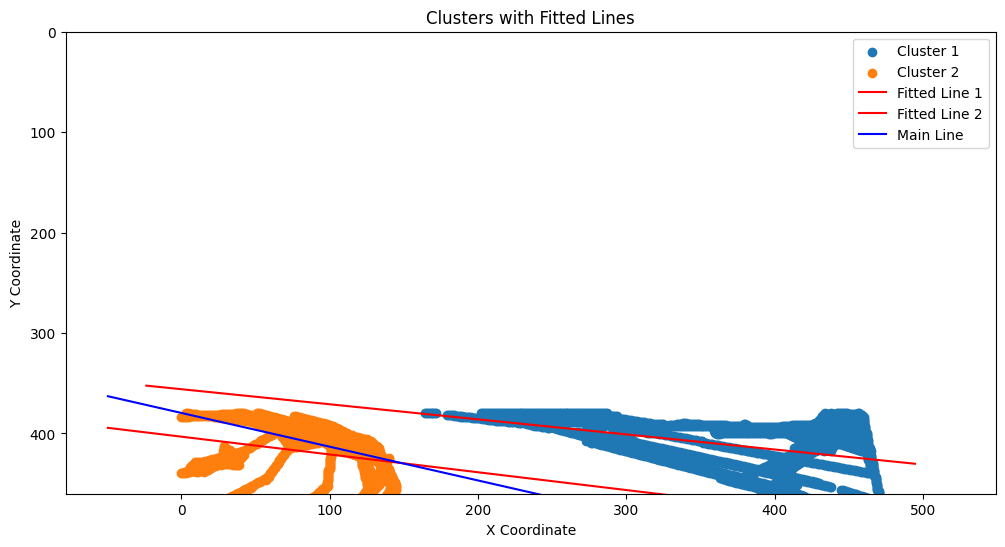}
    \caption{}
    \label{subfig1c_t2}
\end{subfigure}\hfill

\begin{subfigure}{.3\textwidth}
     \includegraphics[width=1.\linewidth]{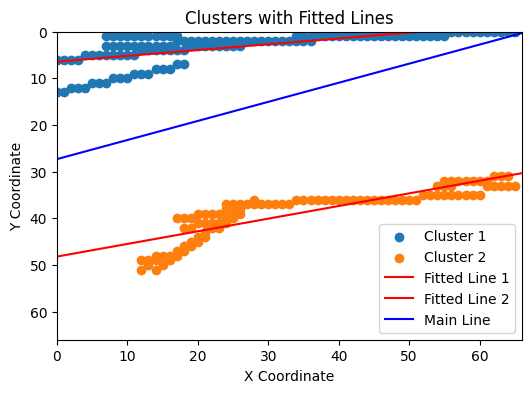}
    \caption{}
    \label{subfig1a_t3}
\end{subfigure}\hfill
\begin{subfigure}{.3\textwidth}
     \includegraphics[width=0.8\linewidth]{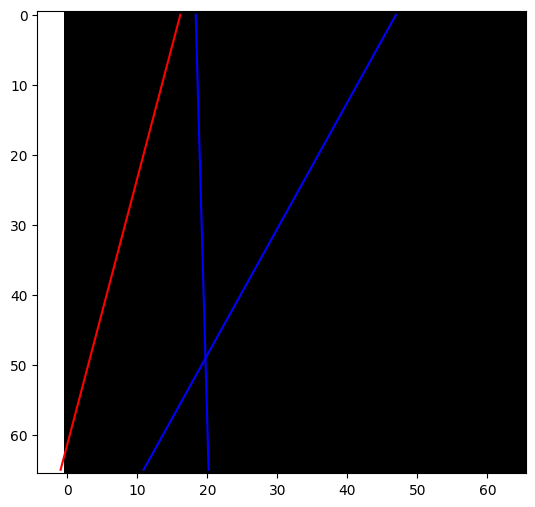}
    \caption{}
    \label{subfig1b_t3}
\end{subfigure}\hfill
\begin{subfigure}{.3\textwidth}
     \includegraphics[width=1.\linewidth]{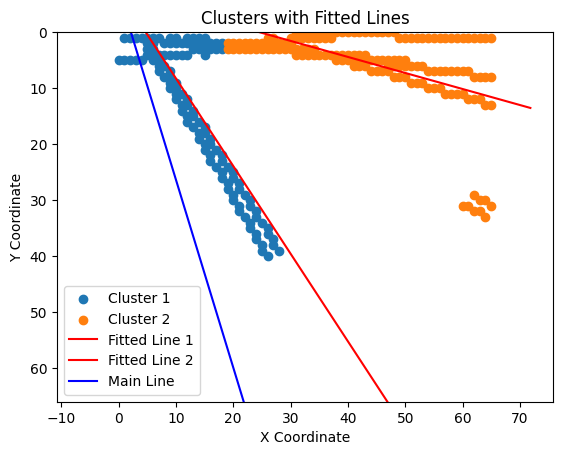}
    \caption{}
    \label{subfig1c_t3}
\end{subfigure}\hfill
%editing here 
\caption{Sample Data Were Analyzed Using Three Clustering Techniques—K-Means Clustering, Image-Splitting, and Spectral Clustering—Applied Consistently Across Track 1 (a-c), Track 2 (d-f), and the CARLA (g-i).}
%Sample data were analyzed using three clustering techniques: Track 1 (a) k-means clustering, (b) image-splitting, and (c) spectral clustering. Similarly, : (d) k-means clustering, (e) image-splitting, and (f) spectral clustering. Similarly, the data from CARLA was processed using the same methods: (g) k-means clustering, (h) image-splitting and (i) spectral clustering}
% \caption{Sample data from Track 1 were analyzed using three clustering techniques: (a) k-means Clustering, (b) Image-splitting, and (c) Spectral Clustering. Similarly, sample data from Track 2 were processed using the same clustering methods: (d) k-means Clustering, (e) Image-splitting technique, and (f) Spectral clustering. Each clustering method has its strengths and limitations. K-means clustering is computationally efficient but tends to struggle with noise and outliers. The image-splitting technique excels at handling variability, delivering consistent results across diverse conditions. Spectral clustering is highly effective in structured, dense environments but requires careful fine-tuning to generalize well. By comparing the performance of these methods across both tracks, valuable insights are gained into their suitability for enhancing autonomous navigation tasks.}
\label{fig5:sampledataforCanny}
\end{figure*}
\subsection{Noise Models}
In QC, noise refers to environmental disturbances that affect quantum state evolution. Kraus operators model this noise by mapping quantum states and ensuring the process is valid through the completeness condition, preserving normalization and unit trace, $\sum_{i=0}^{N} E_{i}^{\dagger} E_{i}=I$ (see Eqs. 11-16).
Here, the Kraus operators \(\{E_i\}\) satisfy the completeness condition, ensuring that the evolved state \(\rho'\) remains a valid quantum state, meaning it is normalized and has a unit trace. Each Kraus operator \(E_i\) acts linearly on the quantum state \(\rho\). The resulting state \(\rho'\) is a weighted sum of these terms, representing the effect of each operator on the initial state. This formalism allows for the description of how quantum operations or channels influence quantum states, accounting for noise and other interactions, $\rho' = \sum_{i=0}^{N} E_{i}^{\dagger} \rho E_{i}$. The I, X, Y, and Z are Pauli operators used in the construction of Kraus operators to represent different types of quantum noise channels, with \( p \) denoting the probability associated with each operation.

\subsubsection{Bitflip}
%The bitflip error occurs when the value of a qubit (quantum bit) is flipped from 0 to 1 or vice versa with a certain probability. The Kraus operator for bit flip is given by,
It occurs when a qubit's state flips between 0 and 1 with a certain probability. The Kraus operators for this error are:
\begin{eqnarray}
E_{0} = \sqrt{(1 - p)} I, \ E_{1} = \sqrt{p} X 
\end{eqnarray}

\subsubsection{Phaseflip}
It changes the qubit's phase, flipping the state from \(\ket{0} + \ket{1}\) to \(\ket{0} - \ket{1}\) or vice versa. The Kraus operators for this error are:
%The phase flip error causes the phase of a qubit to flip. Specifically, a phase flip error transforms the state of a qubit from $\ket{0} + \ket{1}$ to $\ket{0} - \ket{1}$, or vice versa. The Kraus operators for the phase flip error channel are given by:

\begin{eqnarray}
E_{0} = \sqrt{(1 - p)} I, \ E_{1} = \sqrt{p} Z. 
\end{eqnarray}

\subsubsection{BitPhase Flip}
It alters both the bit value and phase of a qubit, transforming \(\ket{0}\) to \(-\ket{1}\) or vice versa. The Kraus operators for this error are:
%The Bitphase flip error is an unintended alteration of both the bit value and the phase of a qubit. This error can transform the state $|0\rangle$ to $-|1\rangle$ or vice versa, resulting in a combination of a bit flip and a phase flip. The Kraus operators for the bit phase flip error channel are given by;

\begin{eqnarray}
E_{0} = \sqrt{1 - p} I, \
E_{1} = \sqrt{p} Y.
\end{eqnarray}

\subsubsection{Depolarizing}
It is a stochastic quantum error where a qubit can randomly rotate about any axis on the Bloch sphere with a certain probability, reflecting the strength of the noise. It describes the effect of an error channel that transforms the state of a qubit as $\rho\longrightarrow (1-p)\rho + pI/2$.
%Depolarizing error is a stochastic error model, which is a very common type of quantum error. It assumes the qubit can randomly rotate about any axis in the Bloch Sphere with a certain probability based on the strength of the noise. More specifically, it describes the effect of an error channel that transforms the state of a qubit as follows:
The Kraus operator for the Depolarising map can be expressed as

\begin{eqnarray}
E_{0} &=& \sqrt{(1 - 3p/4)} I, \
E_{1} = \sqrt{p/4} Z, \nonumber\\
E_{2} &=& \sqrt{p/4} X, \
E_{3} = \sqrt{p/4} Y. 
\end{eqnarray}

\subsubsection{Amplitude Damping}
It is a common error in quantum systems due to energy dissipation. The Kraus operators for this error are:
%Amplitude damping is an important error and usually occurs in quantum systems where the qubits are susceptible to energy dissipation. For amplitude damping, the Kraus operators are given by;

\begin{eqnarray}
E_{0}=\sqrt{p}\ket{0}\bra{1},\
E_{1}=\begin{bmatrix}1 & 0 \\0 & \sqrt{1-p}\end{bmatrix} .   
\end{eqnarray} 

\subsubsection{Phase damping}
It causes the loss of phase information, degrading quantum algorithm performance. The Kraus operators for this error are:
%Phase damping causes the loss of phase information. Phase damping errors can significantly degrade the performance of quantum algorithms and computations. The Kraus operator is given by,

\begin{eqnarray}
    E_{0}=\sqrt{(1-p)}Z, \
    E_{1}=\sqrt{p}\ket{0}\bra{0}, \
    E_{2}=\sqrt{p}\ket{1}\bra{1}   
\end{eqnarray}

\section{Experimental Results \label{SecIII}}
\subsection{Settings and Hyperparameters}
The experiment is conducted in the QASM simulator on the Qiskit platform for 1024 shots, with preprocessing done using OpenCV, including downsampling, Canny function, Gaussian blur, and median filter. The image is downsampled by a factor of 4 and subdivided into $79\times 79$ submatrices for shadow detection. The shadow detection algorithm is applied, followed by 30 iterations of the median filter and Gaussian blur with a kernel size of \((5, 5)\) and a standard deviation of 0. Edge detection uses Canny with thresholds of 50 and 175. The ROI parameters for our algorithm are chosen as $[(0, height), (80, 380), (380, 380), (width, height)]$. The Hough transform uses a pixel resolution of 1, an angular resolution of \(\pi/180\), an accumulator threshold of 8, a minimum line length of 2 pixels, and a maximum gap of 25 pixels. After applying clustering techniques, k-means (Figs. 4a, 4d, 4g) and spectral clustering (Figs. 4c, 4f, 4i) separated data into two groups, shown in orange and blue. Fitted lines highlight the alignment of clustered data points. The image-splitting technique (Figs. 4b, 4e, 4h) divides the input into two halves, fits lines separately, and draws a middle line to represent central alignment, which is used for final predictions. For shadow detection, the classification conditions in the \(UU^\dagger\) method are set to 0.75 and 0.97 for two datasets. Direction prediction is based on the probability differences between two quantum circuits: less than 0.2 predicts ``straight," a higher first-circuit probability predicts ``right", and otherwise ``left". In our model, four parameters are trained for the VQC and one for the quantum shadow detection.

\subsection{Dataset}\label{QVP:Sec2}
%The shadow detection images are captured using a phone camera and sourced from platforms like Pinterest and the data was collected from Kaggle \cite{barui2023novelapproachthresholdquantum}.
For the road dataset, the Udacity driving simulator is used \cite{self-driving-car-sim}. Fig. \ref{fig4:sampledataformodel} shows a subset of modified sample images to fit the input criteria obtained from the Udacity simulator, which provides two tracks: a simpler one (Track 1) and a more complex one (Track 2). Data is collected from both tracks, creating two datasets. The simulator captures frames from three front-facing ``cameras" along with key driving metrics like throttle, speed, and steering angle. These camera frames are the primary input for the model, which predicts steering angles within the range of \([-1, 1]\). The simulator, built in Unity, allows customizable resolution and control settings and saves recorded data in a designated folder for dataset preparation and model training. Additionally, the model was tested using a dataset generated from the CARLA simulator, collected from Kaggle \cite{booni_carla_steering_dataset}. The CARLA dataset, featuring roads affected by rain, demonstrated the model's ability to perform reliably in challenging environmental conditions.

\begin{figure*}
\centering
\begin{subfigure}{.5\textwidth}
    \includegraphics[width=\linewidth]{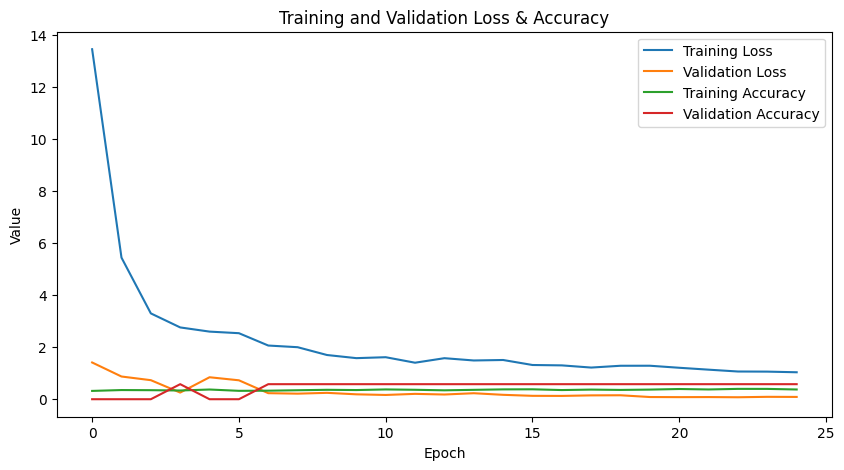}
    \caption{}
    \label{subfig1avalidation_cl}
\end{subfigure}\hfill
\begin{subfigure}{.5\textwidth}
    \includegraphics[width=\linewidth]{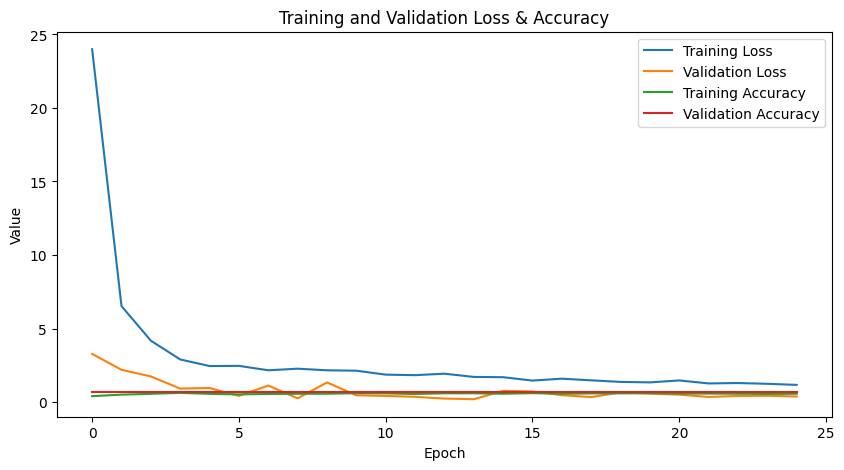}
    \caption{}
    \label{subfig1bvalidatiom}
\end{subfigure}\hfill
\caption{Loss and Accuracy for DNN Training Over 25 Epochs, With Each Epoch Consisting of 5,280 Samples, for (a) Track 1 and (b) Track 2.}
\label{fig4:sampledata_Dataset_1}
\end{figure*}

\begin{figure}
\centering
\begin{subfigure}{.45\textwidth}
    \includegraphics[width=\linewidth]{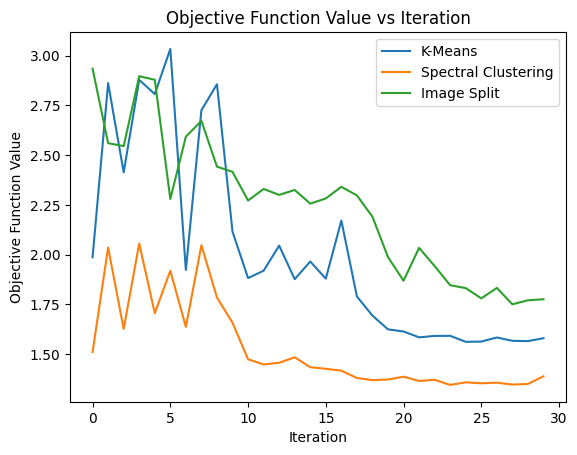}
    \caption{}
    \label{subfig1avalidation}
\end{subfigure}\hfill
\begin{subfigure}{.49\textwidth}
    \includegraphics[width=\linewidth]{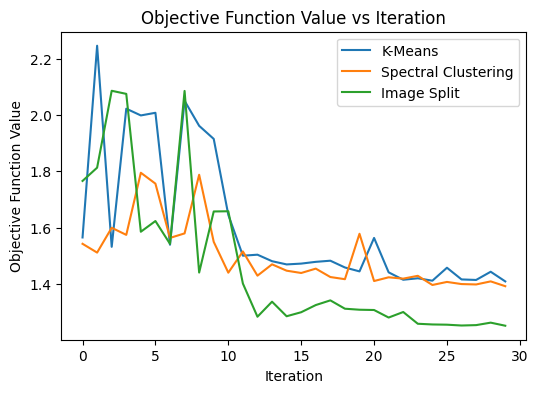}
    \caption{}
    \label{subfig2avalidation}
\end{subfigure}
\begin{subfigure}{.49\textwidth}
    \includegraphics[width=\linewidth]{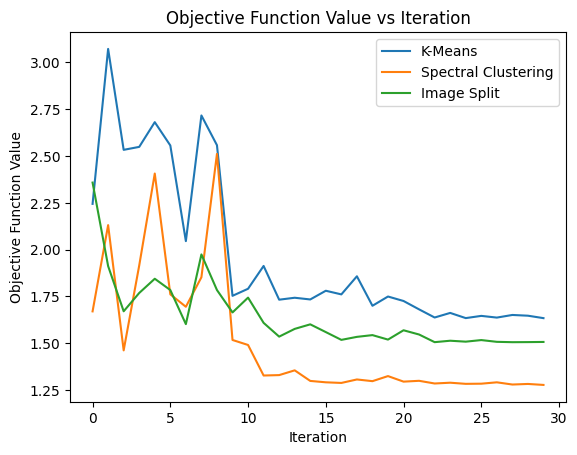}
    \caption{}
    \label{subfig3avalidation}
\end{subfigure}
\caption{The VQC Objective Value Across Iterations for the K-Means, Spectral Clustering, and Image-Splitting: (a) Track 1, (b) Track 2 and (c) CARLA.
\label{fig5:VQC results_1}}
\end{figure}

\begin{figure*}
\centering
\begin{subfigure}{.45\textwidth}
    \includegraphics[width=\linewidth]{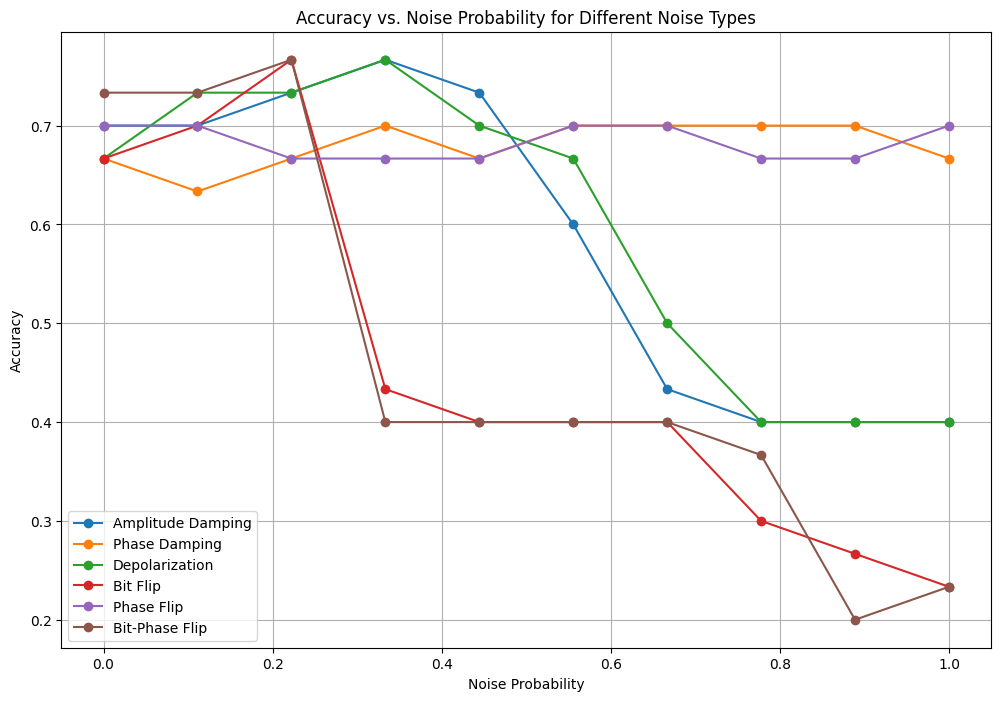}
    \caption{}
    \label{Noisemodeluu1}
\end{subfigure}\hfill
\begin{subfigure}{.45\textwidth}
    \includegraphics[width=\linewidth]{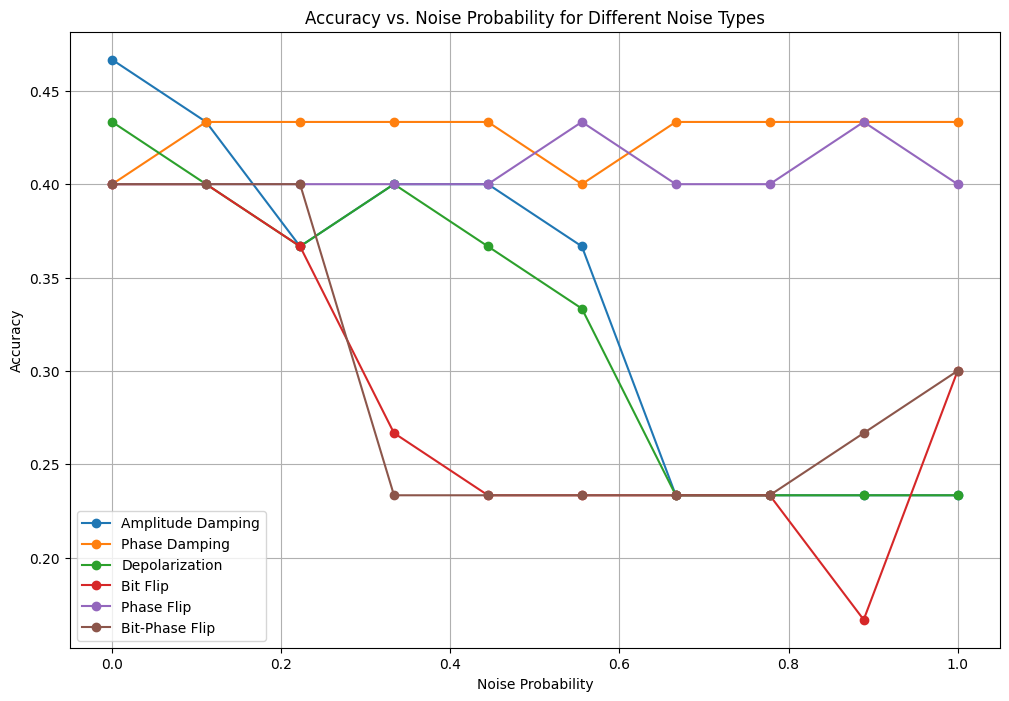}
    \caption{}
    \label{Noisemodeluu2}
\end{subfigure}\hfill
\begin{subfigure}{.45\textwidth}
    \includegraphics[width=\linewidth]{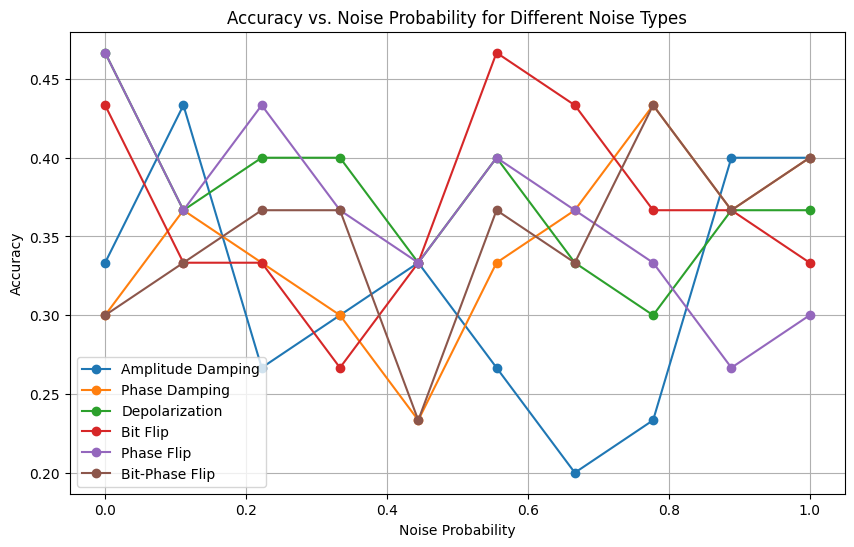}
    \caption{}
    \label{Noisemodelvqc1}
\end{subfigure}\hfill
\begin{subfigure}{.45\textwidth}
    \includegraphics[width=\linewidth]{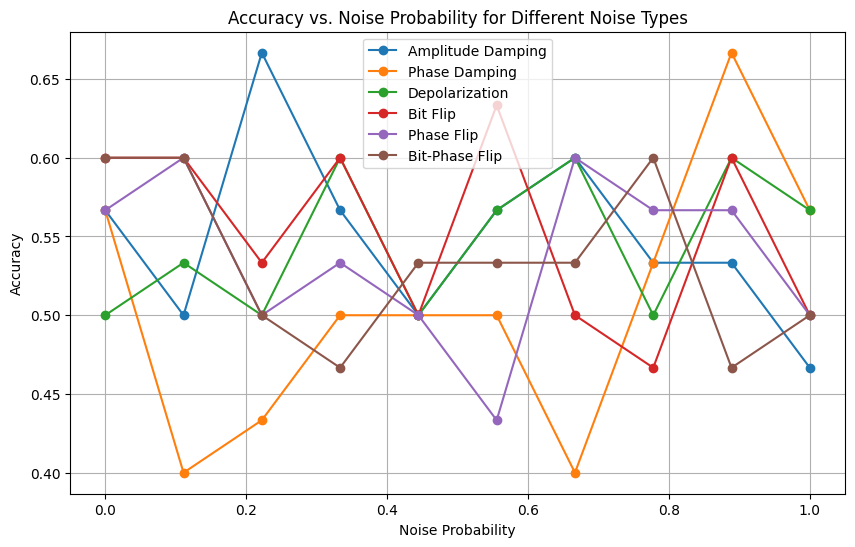}
    \caption{}
    \label{Noisemodelvqc2}
\end{subfigure}\hfill
\begin{subfigure}{.45\textwidth}
    \includegraphics[width=\linewidth]{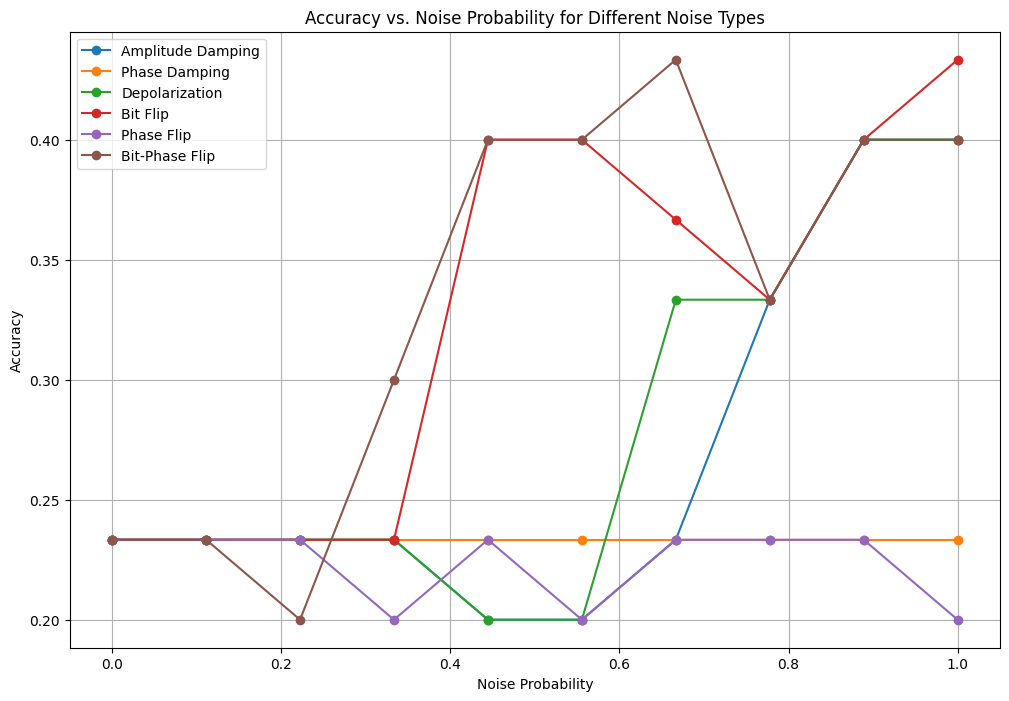}
    \caption{}
    \label{Noisemodeluudagger_carla}
\end{subfigure}\hfill
\begin{subfigure}{.45\textwidth}
    \includegraphics[width=\linewidth]{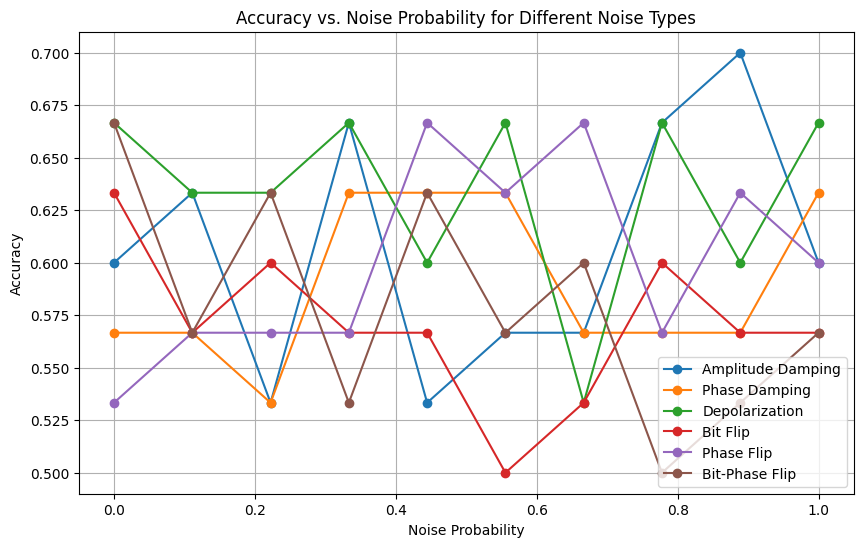}
    \caption{}
    \label{Noisemodelvqc_carla}
\end{subfigure}\hfill
\caption{Performance of UU$^{\dagger}$ and VQC Methods Across Various Noise Types: Track 1 (a, c), Track 2 (b, d), and CARLA (e, f).}
\label{Noise Models}
\end{figure*}

\subsection{Results of Track-1}
In the classical study, a self-driving car system was trained using a DNN that integrated a pre-trained ResNet-50 as its initial layer. The model was trained for 25 epochs with eight batches, processing 46,560 samples. By the final epoch, the training data accuracy peaked at 0.38, while validation accuracy reached 0.58, indicating the model's ability to generalize unseen data as shown in Fig. \ref{fig4:sampledata_Dataset_1}. The model, comprising 26,870,183 parameters, showed decreasing training loss and convergence by the 25th epoch. For the \( UU^\dagger \), the QDCNN algorithm was tested using three variations of the \( UU^\dagger \) method post-Hough transformation. The image-split-\( UU^\dagger \) technique achieved the highest accuracy at 0.7, while k-means-\( UU^\dagger \) and spectral clustering-\( UU^\dagger \) yielded 0.36 and 0.33 accuracy, respectively. For the VQC models, the image-split-VQC method achieved 0.33 accuracy, k-means-VQC scored 0.36, and spectral clustering-VQC performed the best with 0.4 accuracy. The VQC models were optimized using the COBYLA optimizer, and accuracy vs noise strength was plotted (Fig. \ref{Noise Models}). Overall, the quantum approach outperformed the classical one, with the image-split-\( UU^\dagger \) method reaching an accuracy of 0.7, while the best classical accuracy was 0.58. A comparison among the above methods in terms of accuracy is shown in Table \ref{tab:my_labl}.

\subsection{Results of Track-2} 
The model was trained for 25 epochs, each with seven batches, processing 46,560 samples. Initially, the model had a training loss of 1.1598 and an accuracy of 0.55, while validation metrics showed a loss of 0.1094 and an accuracy of 0.705. After fluctuations, the final training loss was 0.7930 with an accuracy of 0.58, and the validation accuracy stabilized at 0.70 with a loss of 0.0473 (see Fig. \ref{fig4:sampledata_Dataset_1}).
Similarly, three \(UU^{\dagger}\) techniques were applied. The image-split-\(UU^{\dagger}\) method achieved the highest accuracy of 0.4, while k-means-\(UU^{\dagger}\) and spectral-clustering-\(UU^{\dagger}\) reached 0.33 and 0.36, respectively.  In the VQC-based models, image-split-VQC achieved 0.6 accuracy, outperforming k-means-VQC and spectral clustering-VQC, which both reached 0.5. While the DNN outperformed QDCNN, the best QDCNN result was from image-split-VQC with 0.6 accuracy (Fig. \ref{fig5:VQC results_1}). It is important to note that proper preprocessing and clustering were crucial in achieving these results. Table \ref{tab:my_labl} provides a comparison of accuracies among the above different methods.

\begin{table}
    \centering
    \begin{tabular}{|c|c|c|c|c|}
    \hline
    
    \textbf{Dataset} & \textbf{Algorithm} & \textbf{Method}  & \textbf{Accuracy}  \\
    
    \hline
       Track - 1 & DNN & & 0.57\\
    \hline
       &QDCNN & Image-Split-$UU^{\dagger}$   &\textbf{0.7} \\
    \hline
         & & Image-Split-VQC   &0.33\\
    \hline
      &  & K-Means-VQC   &0.36 \\
    \hline
      & & Spectral-Clustering-VQC   &0.4 \\

    \hline
      & & K-Means-$UU^{\dagger}$   &0.36 \\
    \hline
    & & Spectral-Clustering-$UU^{\dagger}$   &0.4 \\
    \hline

       Track - 2 & DNN & & 0.70 \\
    \hline
        & QDCNN & Image-Split-$UU^{\dagger}$& 0.4 \\
    \hline
        &  & Image-Split-VQC   &\textbf{0.6}\\
    \hline
       & & K-Means-VQC   &0.5 \\
    \hline
      & & Spectral-Clustering-VQC &0.5 \\
    \hline
      & & K-Means-$UU^{\dagger}$   &0.36 \\
    \hline
     & & Spectral-Clustering-$UU^{\dagger}$   &0.33 \\
    \hline
    \end{tabular}
    \caption{The Accuracy Comparison of Different Algorithms on the Track 1 and Track 2 Datasets.}
    \label{tab:my_labl}
\end{table}

\begin{table}[h!]
\centering
\begin{tabular}{|c|c|}
\hline
\textbf{Algorithm}       & \textbf{Accuracy / Success Rate in \%}       \\ \hline
{DQN-PF \cite{REDA2024104630}}    & 88.62 \\ \hline
{RL-MOHH \cite{REDA2024104630}}   & 80                          \\ \hline
{\textbf{UU$^{\dagger}$}}& \textbf{70.56} \\
\hline
{NoveL CNN based Model \cite{abc}}& 99.56\\
\hline
\end{tabular}

\caption{
Comparison of the Proposed Algorithm with Existing Approaches on Similar Datasets and Their Accuracy/Success Rates}
\label{comparison}
\end{table}

\subsection{Results of Carla Dataset} % Modification left
% For the CARLA dataset, Similar, three \(UU^{\dagger}\) techniques were applied. The image-split-\(UU^{\dagger}\) method achieved the highest accuracy of 40\%, while k-means-\(UU^{\dagger}\) and spectral-clustering-\(UU^{\dagger}\) reached 33\% and 36\%, respectively.  In the VQC-based models, image-split-VQC achieved 60\% accuracy, outperforming k-means-VQC and spectral clustering-VQC, which both reached 50\%. The best QDCNN result was from image-split-VQC with 60\% accuracy (Fig. \ref{fig5:VQC results_1}). It is important to note that proper preprocessing and clustering were crucial in achieving these results. Table \ref{tab:my_labl carla} provides a comparison of accuracies among the above different methods.

For the CARLA dataset, multiple \(UU^{\dagger}\) and VQC-based techniques were applied to evaluate the performance of QDCNN variants. Among the VQC-based models, spectral-clustering-VQC achieved the highest accuracy of 0.6, followed by k-means-VQC with 0.5 and image-split-VQC with 0.41. In contrast, the \(UU^{\dagger}\)
based methods exhibited lower accuracy levels, with image-split-\(UU^{\dagger}\) reaching 0.23, while k-means-\(UU^{\dagger}\) and spectral-clustering- \(UU^{\dagger}\) both achieved 0.2. The results indicate that spectral-clustering-VQC is the most effective approach for the CARLA dataset, highlighting the importance of selecting the appropriate clustering technique as shown in Table \ref{tab:my_label_carla}. %Additionally, proper preprocessing and clustering methods significantly influenced the final accuracy. Table} \ref{tab:my_label_carla} \hl{provides a detailed comparison of the accuracies achieved by different algorithms and clustering techniques.}

\begin{table}
    \centering
    \begin{tabular}{|c|c|}
    \hline
    \textbf{Method}  & \textbf{Accuracy}  \\
    \hline
    Spectral-Clustering-VQC & \textbf{0.6} \\
    \hline
     Image-Split-VQC   & 0.41\\
    \hline
    K-Means-VQC   & 0.5 \\
    \hline
    Image-Split-$UU^{\dagger}$& 0.23 \\
    \hline
    K-Means-$UU^{\dagger}$   & 0.2 \\
    \hline
    Spectral-Clustering-$UU^{\dagger}$   & 0.2 \\
    \hline
    \end{tabular}
    \caption{ The Accuracy Comparison of Different QDCNN Methods on the CARLA Dataset.}
    \label{tab:my_label_carla}
\end{table}

\subsection{Effect of Noise Models}
The study evaluates the performance of the UU$^{\dagger}$ and VQC methods under six types of quantum noise across Tracks 1 and 2 and the CARLA dataset, as shown in Fig. \ref{Noise Models}. The UU$^{\dagger}$ method performed better on Track 1 ((Fig. \ref{Noisemodeluu1}), with a peak accuracy of 0.766 under amplitude damping at a noise probability of 0.22 and stable performance under phase damping and depolarization with accuracy 0.633 and 0.733). However, it experienced sharp declines under bit-flip and bit-phase flip noises, dropping to 0.2 and 0.233, respectively. On Track 2, its performance was lower, with amplitude damping starting at 0.466 and dropping to 0.233, while bit flip and bit-phase flip noises caused significant drops to 0.166 (Fig. \ref{Noisemodeluu2}). For the CARLA dataset, the UU$^{\dagger}$ method struggled, maintaining a low accuracy of 0.233 across most noise types, with minor peaks of 0.4 under amplitude damping and 0.433 under bit flip noise (Fig. \ref{Noisemodeluudagger_carla}).
The VQC method on Track 1 peaked accuracy at 0.433 under amplitude damping and phase damping at a noise probability of 0.11 and 0.78. Bit-flip noise showed fluctuations, peaking at 0.466 at a noise probability of 0.55 and declining at higher probabilities (Fig. \ref{Noisemodelvqc1}). On Track 2, it demonstrated robustness, peaking at 0.666 under amplitude damping and phase damping at a noise probability of 0.22 and 0.88, with consistent accuracy between 0.5 and 0.6 across other noise types (Fig. \ref{Noisemodelvqc2}). The VQC method achieved better performance for the CARLA dataset, peaking at 0.7 under amplitude damping at a noise probability of 0.88. Phase damping showed stable performance, peaking at 0.633 and fluctuating between 0.533 and 0.666, peaking at 0.666 under depolarization while maintaining stability between 0.566 and 0.633 under bit flip and bit-phase flip noises (Fig. \ref{Noisemodelvqc_carla}). Overall, the VQC method consistently outperformed the UU$^{\dagger}$ method, demonstrating better robustness and adaptability to quantum noise, particularly under amplitude and phase damping.

\begin{table}
    \centering
    \begin{tabular}{|c|c|}
    \hline
        \textbf{Algorithm} & \textbf{Time Taken in Seconds}  \\
        \hline
        IBT &0.03\\
        \hline
        CBS & 1.47\\
        \hline
        LBP & 2.05 \\
        \hline
        \textbf{QDCNN} & \textbf{0.0049352} \\
        \hline
    \end{tabular}
    \caption{Time Comparison of Different Algorithms for Shadow Detection}
    \label{tabfortimecomp}
\end{table}
\section{Discussion and Conclusion \label{SecIV}}
This study presents a QDCNN optimized for self-driving cars, incorporating the \(UU^{\dagger}\) method and other quantum algorithms. A key feature is shadow detection using the \(UU^{\dagger}\) method, with shadow removal aiding road detection. The quantum approach, supported by preprocessing, shows effective decision-making, achieving peak accuracy of 0.7 in initial dataset trials with the \(UU^{\dagger}\) method and 0.6 on Track 2 using the VQC. Conversely, a classical DNN leveraging ResNet-50 trained over 25 epochs with 46,560 samples achieves a peak training accuracy of 0.38 and a validation accuracy stabilizing at 0.58. On Track 2, the DNN outperforms the quantum model, achieving 0.70 validation accuracy. A comparison among existing work on similar datasets is shown in Table \ref{comparison}. In terms of time efficiency, several classical algorithms were compared (Table \ref{tabfortimecomp}): Intensity-based Thresholding (IBT) took 0.03 seconds, Chromaticity-based Shadow detection (CBS) took 1.47 seconds, and the Local Binary Pattern (LBP) technique took 2.05 seconds. In contrast, our quantum algorithm, operating on 224,676 elements based on photonic quantum gate operation times \cite{PhysRevLett.133.090601}, which ranges from picoseconds to femtoseconds assuming 1000 runs, was significantly faster, taking only 0.0049352 seconds.
The proposed QDCNN is resource-efficient, using grayscale images and requiring only one qubit and two gates, making it ideal for integration into UAVs and self-driving cars with limited resources. It has the potential to significantly enhance safety and reliability in autonomous transportation systems by improving the accuracy of key tasks such as shadow detection and decision-making in dynamic environments. This increased accuracy in detecting obstacles and road boundaries reduces the risk of accidents caused by environmental factors like shadows or poor visibility. Moreover, the robustness of the quantum models, when tested under various noise conditions, demonstrates their reliability in real-world scenarios where external disturbances are common. The ability to integrate advanced preprocessing techniques with quantum algorithms allows for more accurate predictions and decision-making, improving the vehicle's responsiveness to changes in its surroundings. By expanding the system to identify pedestrians and vehicles, the overall situational awareness of autonomous cars will be enhanced, making transportation systems safer and more dependable. Incorporating these quantum methods into autonomous driving architectures will not only increase the system’s capacity to handle complex tasks but also improve reliability in unpredictable environments, leading to safer, more efficient transportation systems.

\ifCLASSOPTIONcaptionsoff
\newpage
\fi

\bibliographystyle{IEEEtran}

\bibliography{IEEE}

\vfill

\end{document}